\documentclass[12pt]{article}

\ifx\pdfoutput\undefined
\usepackage[dvips,bookmarks]{hyperref}
\else
\usepackage{hyperref}
\fi
\hypersetup{colorlinks=false,bookmarksopen,bookmarksnumbered,citecolor=blue,
   pdfstartview=FitH}

\usepackage[dvips]{graphicx}
\usepackage{latexsym}
\usepackage{amssymb,amsfonts,amsmath}
\usepackage{graphicx} 
\usepackage{indentfirst}

 \usepackage{bbm}

\parskip=\medskipamount

\newcommand {\cC}{{\cal C}}
\newcommand {\cD}{{\cal D}}
\newcommand {\cE}{{\cal E}}

\newcommand {\cG}{{\cal G}}

\newcommand {\cL}{{\cal L}}
\newcommand {\cM}{{\cal M}}
\newcommand {\cN}{{\cal N}}

\newcommand {\cQ}{{\cal Q}}
\newcommand {\cR}{{\cal R}}
\newcommand {\cS}{{\cal S}}
\newcommand {\cT}{{\cal T}}

\newcommand {\cV}{{\cal V}}
\newcommand {\cW}{{\cal W}}
\newcommand {\cX}{{\cal X}}

\def\a{\alpha}
\def \bi{\bibitem}

\def\b{\beta}

\def\d{\delta}

\def\g{\gamma}
\def\G{\Gamma}

\def\k{\kappa}
\def\l{\lambda}
\def\m{\mu}
\def\n{\nu}

\def\q{\theta}

\newcommand{\qb}{{\bar{\theta}}}

\def\s{\sigma}

\def\z{\zeta}

\def\F{\Phi}

\def\L{\Lambda}
\def\O{\Omega}

\def\S{\Sigma}
\def\U{\Upsilon}

\def\rd{{\rm d}}
\def\ri{{\rm i}}
\def\re{{\rm e}}

\newcommand{\ad}{{\dot{\alpha}}}                           
\newcommand{\bd}{{\dot{\beta}}}                            
\newcommand{\ve}{\varepsilon}                            

\newcommand{\pa}{\partial}                           
\newcommand{\hf}{\frac12}

\newcommand{\be}{\begin{equation}}
\newcommand{\ee}{\end{equation}}
\newcommand{\bea}{\begin{eqnarray}}
\newcommand{\eea}{\end{eqnarray}}
\newcommand{\non}{\nonumber}
\newcommand{\1}{{\underline{1}}}
\newcommand{\2}{{\underline{2}}}

\def\dt#1{{\buildrel {\hbox{\LARGE .}} \over {#1}}}    

\newcommand{\bm}[1]{\mbox{\boldmath$#1$}}

\def\double #1{#1{\hbox{\kern-2pt $#1$}}}

\newcommand{\hm}{{\hat{m}}}

\newcommand{\ha}{{\hat{a}}}
\newcommand{\hb}{{\hat{b}}}
\newcommand{\hc}{{\hat{c}}}
\newcommand{\hd}{{\hat{d}}}
\newcommand{\he}{{\hat{e}}}

\newcommand{\hal}{{\hat{\a}}}
\newcommand{\hbe}{{\hat{\b}}}
\newcommand{\hga}{{\hat{\g}}}
\newcommand{\hde}{{\hat{\d}}}

\begin{document}

\begin{titlepage}

\begin{flushright}
April, 2008\\
\end{flushright}
\vspace{5mm}

\begin{center}
{\Large \bf  Conformally flat  supergeometry in five dimensions}\\ 
\end{center}

\begin{center}

{\large  
Sergei M. Kuzenko\footnote{{kuzenko@cyllene.uwa.edu.au}}
and 
Gabriele Tartaglino-Mazzucchelli\footnote{gtm@cyllene.uwa.edu.au}
} \\
\vspace{5mm}

\footnotesize{
{\it School of Physics M013, The University of Western Australia\\
35 Stirling Highway, Crawley W.A. 6009, Australia}}  
~\\

\vspace{2mm}

\end{center}
\vspace{5mm}

\begin{abstract}
\baselineskip=14pt
Using the superspace formulation for the 5D $\cN=1$ Weyl supermultiplet 
developed in  arXiv:0802.3953, we elaborate the concept of conformally 
flat superspace in five dimensions. For a large family of supersymmetric 
theories (including sigma-models and Yang-Mills theories)
in the conformally flat superspace, we describe an explicit
procedure to formulate their dynamics in terms of rigid 4D $\cN=1$ superfields.
The case of 5D $\cN=1$ anti-de Sitter
superspace is discussed as an example.
\end{abstract}
\vspace{1cm}

\vfill
\end{titlepage}

\newpage
\renewcommand{\thefootnote}{\arabic{footnote}}
\setcounter{footnote}{0}

In the context of 
the two-brane Randall-Sundrum scenario \cite{RS}
and its supersymmetric extensions
\cite{ABN,GP,FLP}, 
it is of  interest
to have a superspace description 
for five-dimensional $\cN=1$ conformally flat supergeometry
that would be similar to that available in the case of four-dimensional 
$\cN=1$ supersymmetry, see, e.g. \cite{BK} for a review.
This is also an interesting problem from the point of view of formal supergravity.
Such a description can be derived using the superspace formulation 
for the Weyl multiplet of 5D $\cN=1$ conformal supergravity \cite{Ohashi,Bergshoeff}, 
which has recently been given in \cite{KT-Msugra3}
(building on  \cite{KT-Msugra1,KT-Msugra2}).
Its elaboration is provided  in the present letter.
The case of 5D $\cN=1$ anti-de Sitter superspace, 
which was studied in \cite{KT-M} from a different perspective, 
is explicitly worked out  as an example.

To start with, it is worth recalling the salient points of
the superspace formulation developed in  \cite{KT-Msugra3}.
Let $z^{\hat{M}}=(x^{\hm},\q^{\hat{\mu}}_i)$
be local bosonic ($x$) and fermionic ($\q$) 
coordinates parametrizing  a curved five-dimensional $\cN=1$  superspace
$\cM^{5|8}$,
where $\hm=0,1,\cdots,4$, $\hat{\mu}=1,\cdots,4$, and  $i=\1,\2$.
Here the Grassmann variables $\q^{\hat{\mu}}_i$
 obey the standard pseudo-Majorana reality condition
$\overline{\q^{\hat{\mu}}_i }= \q_{\hat{\mu}}^i =\ve_{\hat{\m} \hat{\n}}\,  \ve^{ij} \, \q^{\hat{\nu}}_j  $
(see the appendix in \cite{KT-Msugra2} for our 5D  notation and conventions).
The tangent-space group
is chosen to be  ${\rm SO}(4,1)\times {\rm SU}(2)$,
and the superspace  covariant derivatives 
$\cD_{\hat{A}} =(\cD_{\hat{a}}, \cD_{\hat{\a}}^i) $
have the form 
\bea
\cD_{\hat{A}}&=&
\cE_{\hat{A}} + \O_{\hat{A}} + \F_{\hat{A}}
~.
\label{CovDev}
\eea
Here $\cE_{\hat{A}}= \cE_{\hat{A}}{}^{\hat{M}}(z) \,\pa_{\hat{M}}$ is the supervielbein, 
with $\pa_{\hat{M}}= \pa/ \pa z^{\hat{M}}$,
\bea
\O_{\hat{A} }= \hf \,\O_{\hat{A}}{}^{\hb\hc}\,M_{\hb\hc}
= \O_{\hat{A}}{}^{\hbe\hga}\,M_{\hbe\hga}~,\qquad 
M_{\ha\hb}=-M_{\hb\ha}~, \quad M_{\hal\hbe}=M_{\hbe\hal}
\eea
is the Lorentz connection, 
\bea
\F_{\hat{A}} = \F^{~\,kl}_{\hat{A}}\,J_{kl}~, \qquad
J_{kl}=J_{lk}
\eea
is the SU(2)-connection. 
The Lorentz generators with vector indices ($M_{\ha\hb}$) and spinor indices
($M_{\hal\hbe}$) are related to each other by the rule:
$M_{\ha\hb}=(\S_{\ha\hb})^{\hal\hbe}M_{\hal\hbe}$ 
(for more details, see the appendix of \cite{KT-Msugra2}).
The generators of ${\rm SO}(4,1)\times {\rm SU}(2)$
act on the covariant derivatives as follows:\footnote{The operation of
(anti)symmetrization of $n$ indices 
is defined to involve a factor $(n!)^{-1}$.}
\bea
{[}J^{kl},\cD_{\hal}^i{]}
= \ve^{i(k} \cD^{l)}_{\hat \a}~,~~~
{[}M_{\hal\hbe},\cD_{\hga}^k{]}
=\ve_{\hga(\hal}\cD^k_{\hbe)}~,~~~
{[}M_{\ha\hb},\cD_{\hc}{]}
=2\eta_{\hc[\ha}\cD_{\hb]}~,
\label{generators}
\eea
where $J^{kl} =\ve^{ki}\ve^{lj} J_{ij}$.

The covariant derivatives obey (anti)commutation relations of the general form 
\bea
{[}\cD_{\hat{A}},\cD_{\hat{B}}\}&=&\cT_{\hat{A}\hat{B}}{}^{\hat{C}}\cD_{\hat{C}}
+\hf \cR_{\hat{A}\hat{B}}{}^{\hat{c}\hat{d}}M_{\hat{c}\hat{d}}
+\cR_{\hat{A}\hat{B}}{}^{kl}J_{kl}
~,
\label{algebra}
\eea
where $\cT_{\hat{A}\hat{B}}{}^{\hat{C}}$ is the torsion, 
$\cR_{\hat{A}\hat{B}}{}^{\hat{c}\hat{d}}$
and $\cR_{\hat{A}\hat{B}}{}^{kl}$ 
the  SO(4,1)  and SU(2) curvature tensors, respectively. 

To describe the  Weyl multiplet of conformal supergravity  \cite{Ohashi,Bergshoeff}, 
the torsion has to obey the constraints \cite{KT-Msugra3}:
\bea
\cT_{\hal}^i{}_{\hbe}^j{}^{\hc}~=~-2\ri\ve^{ij}(\Gamma^{\hc})_{\hal\hbe},  \qquad
\cT_{\hal}^i{}_{\hbe}^j{}^{\hga}_k~=~
\cT^i_{\hal}{}_{\hb}{}^{\hc}~
=~0,  \qquad  
\cT_{\ha\hb}{}^{\hc}~=~\cT_{\ha\hbe (j}{}^{\hbe}{}_{k)}~=~0 ~.
\label{constr}
\eea
With  the constraints introduced,  it can be shown that
the torsion and the curvature tensors in (\ref{algebra}) 
are expressed in terms of a small number of 
dimension-1 tensor superfields, $\cS^{ij}$, $\cX_{\ha\hb}$, 
$\cN_{\ha\hb}$ and $\cC_\ha{}^{ij}$, and their covariant derivatives,  
with the symmetry properties:
\bea
\cS^{ij}=\cS^{ji}~,\qquad \cX_{\ha\hb}=-\cX_{\hb\ha}~,\qquad \cN_{\ha\hb}=-\cN_{\hb\ha}~,
\qquad \cC_\ha{}^{ij}=\cC_\ha{}^{ji}~.
\eea
Their reality properties are
\be
\overline{\cS^{ij} } =\cS_{ij}~, \qquad 
\overline{\cX_{\ha\hb}} =\cX_{\ha\hb}~, \qquad 
\overline{\cN_{\ha\hb}} =\cN_{\ha\hb}~, \qquad
\overline{\cC_\ha{}^{ij} }=\cC_{\ha ij }~.
\ee
The covariant derivatives obey the (anti)commutation relations  \cite{KT-Msugra3}:
\begin{subequations}
\bea
\big\{ \cD_{\hal}^i , \cD_{\hbe}^j \big\} &=&-2 \ri \,\ve^{ij}\cD_{\hal\hbe}
-\ri \,\ve_{\hal\hbe}\ve^{ij}\cX^{\hc\hd}M_{\hc\hd}
+{\ri\over 4} \ve^{ij}\ve^{\ha\hb\hc\hd\he}(\G_\ha)_{\hal\hbe}\cN_{\hb\hc}M_{\hd\he}
\non\\
&&
-{\ri\over 2}\ve^{\ha\hb\hc\hd\he}(\S_{\ha\hb})_{\hal\hbe}\cC_{\hc}{}^{ij}M_{\hd\he}
+4\ri \,\cS^{ij}M_{\hal\hbe}
+3\ri \, \ve_{\hal\hbe}\ve^{ij}\cS^{kl}J_{kl}
\non\\
&&
-\ri \, \ve^{ij}\cC_{\hal\hbe}{}^{kl}J_{kl}
-4\ri\Big(\cX_{\hal\hbe}+\cN_{\hal\hbe}\Big)J^{ij}
~,
\label{covDev2spinor-} \\
{[}\cD_\ha,\cD_{\hbe}^j{]}&=&
{1\over 2} \Big(
(\Gamma_{\hat{a}})_{\hbe}{}^{\hga}\cS^j{}_k
- \cX_{\ha\hb}(\Gamma^{\hat{b}})_{\hbe}{}^{\hga} \d^j_k
-{1\over 4}\,\ve_{\ha\hb\hc\hd\he}\cN^{\hd\he}(\Sigma^{\hb\hc})_{\hbe}{}^{\hga}
\d^j_k
+ (\S_\ha{}^{\hb})_{\hbe}{}^{\hga}\cC_\hb{}^j{}_k
\Big)
\cD_{\hga}^k
\non\\
&&
\qquad \qquad 
~+~\mbox{curvature terms}~.
\label{covDev2spinor-2}
\eea
\end{subequations}
The dimension-1 components of the torsion, 
 $\cS^{ij}$, $\cX_{\ha\hb}$, 
$\cN_{\ha\hb}$ and $\cC_\ha{}^{ij}$, 
enjoy some additional differential constraints 
implied by the Bianchi identities \cite{KT-Msugra3}.

Let $D_{\hat{A}} =(D_{\hat{a}}, D_{\hat{\a}}^i) $
be another set of covariant derivatives satisfying the constraints (\ref{constr}), 
with  $S^{ij}$, $X_{\ha\hb}$, $N_{\ha\hb}$ and $C_\ha{}^{ij}$ 
being the corresponding dimension-1 components of the torsion.
The supergeometries, which are associated with 
$\cD_{\hat{A}} $ and $D_{\hat{A}} $, describe the same Weyl multiplet if they 
are related by a super-Weyl transformation\footnote{In \cite{KT-Msugra3},
only the infinitesimal    
super-Weyl transformation was explicitly given.}
\cite{KT-Msugra3} of the form:
\begin{subequations}
\bea
\cD_\hal^i&=&
\re^{\s}\Big(D_\hal^i+4(D^{\hbe i}\s)M_{\hal\hbe}-6(D_{\hal j}\s)J^{ij}\Big)~,
\label{D_hal-ConfFlat}
\\
\cD_{\ha}&=&
\re^{2\s}\Big{(}
D_\ha
+\ri(\G_\ha)^{\hga\hde}(D_\hga^k\s)D_{\hde k}
-2(D^\hb\s)M_{\ha\hb}
+{\ri\over 4}(\G_\ha)^{\hga\hde}(D_{\hga}^{k}D_{\hde}^{l}\s)J_{kl}
\non\\
&&~~~~~~
+{\ri\over 2}\ve_{\ha\hb\hc\hd\he}(\S^{\hb\hc})_{\hga\hde}(D^{\hga k}\s)(D^{\hde}_k\s)M^{\hd\he}
+{5\ri\over 2}(\G_\ha)^{\hga\hde}(D_{\hga}^k\s)(D_{\hde}^l\s)J_{kl}
\Big{)}
~.~~~~~~
\label{D_a-ConfFlat-2}
\eea
\end{subequations}
The components of the torsion are related as follows: 
\begin{subequations}
\bea
\cX_{\hc\hd}&=&
\re^{2\s}\Big{(}
X{}_{\hc\hd}
-{\ri\over 2}(\S_{\hc\hd})_{\hga\hde}(D^{\hga k}D^{\hde}_k\s)
-3\ri(\S_{\hc\hd})_{\hga\hde}(D^{\hga k}\s)(D^{\hde}_k\s)
\Big{)}
~,~~~
\label{F-ConfFlat-2}
\\
\cN_{\hc\hd}&=&
\re^{2\s}\Big{(}
N{}_{\hc\hd}
-\ri(\S_{\hc\hd})_{\hga\hde}(D^{\hga k}D^{\hde}_{k}\s)
-6\ri(\S_{\hc\hd})_{\hga\hde}(D^{\hga k}\s)(D^{\hde}_{k}\s)
\Big{)}
~,
\label{N-ConfFlat-2}
\\
\cC_{\ha}{}^{jk}
&=&
\re^{2\s}\Big{(}
C{}_{\ha}{}^{jk}
+\ri(\G_\ha)^{\hal\hbe}(D_{\hal}^{(j}D_{\hbe}^{k)}\s)
-2\ri(\G_\ha)^{\hal\hbe}(D_{\hal}^{(j}\s)(D_{\hbe}^{k)}\s)
\Big{)}
\label{C-ConfFlat-2}
~,\\
\cS^{ij}&=&
\re^{2\s}\Big{(}
S{}^{ij}
+{\ri\over 2}(D^{\hga(i}D_{\hga}^{j)}\s)
-3\ri(D^{\hga (i}\s)(D_{\hga}^{j)}\s)
\Big{)}
~.
\label{S-ConfFlat-2}
\eea
\end{subequations}
Consider the super-Weyl tensor \cite{KT-Msugra3} 
\be
\cW_{\ha \hb} := \cX_{\ha\hb} -\hf  \cN_{\ha\hb}~.
\ee
It follows from eqs. (\ref{F-ConfFlat-2}) and (\ref{N-ConfFlat-2}) 
that it transforms homogeneously, 
\be
\cW_{\ha \hb} = \re^{2\s}\, W_{\ha \hb} ~.
\ee
If the supergeometry  $D_{\hat{A}} $ is such that its super-Weyl tensor vanishes,
$W_{\ha \hb} =0$, the same property holds for the supergeometry  $\cD_{\hat{A}} $.
If the supergeometry  $D_{\hat{A}} $ is flat, the supergeometry  $\cD_{\hat{A}} $
will be called {\it conformally flat}.

Suppose that the two supergeometries under consideration are 
such that\footnote{As observed in  \cite{KT-Msugra3}, the super-Weyl gauge 
freedom can always be used to choose the gauge $ \cC{}_{\ha}{}^{ij}=0$.}
\be
\cC{}_{\ha}{}^{ij}= C{}_{\ha}{}^{ij}=0~.
\label{C=0}
\ee
Then, it follows from  (\ref{C-ConfFlat-2}) that the parameter $\s$ is constrained. 
The relevant constraint can be expressed in the form:
\bea
D_{\hal}^{(i}D_{\hbe}^{j)}W_0
-{1\over 4}\ve_{\hal\hbe} D^{\hga(i} D_{\hga}^{j)} W_0 =0~,
\qquad W_0:=\re^{-2\s}~.
\label{W-BI-2}
\eea
This is the equation for the {\it field strength of an Abelian 
vector multiplet}. In what follows, we will assume the fulfillment of (\ref{C=0}).

More generally, consider an arbitrary non-Abelian vector multiplet. 
Its field strength $\cW$ obeys the constraint 
\bea
\cD_{\hal}^{(i}\cD_{\hbe}^{j)}\cW
-{1\over 4}\ve_{\hal\hbe}\cD^{\hga(i}\cD_{\hga}^{j)} \cW =0
~~~~~~~
\label{W-BI}
\eea
and possesses the super-Weyl transformation
\bea
\cW={\rm e}^{2\s} W~.
\label{super-Weyl-W}
\eea
Associated with the vector multiplet is 
the  composite superfield  \cite{KT-Msugra3} 
\bea
\cG^{ij}:= {\rm tr} \,\Big\{ 
\ri\,\cD^{\hal (i}\cW\cD_\hal^{j)} \cW+{\ri\over 2} \cW \cD^{ij} \cW-2\cS^{ij}\cW^2 \Big\}~,
\qquad \cD^{ij}:= \cD^{\hal (i} \cD_\hal^{j)}  ~,
\label{Gij}
\eea
which enjoys the equation
\bea
\cD^{(i}_\hal \cG^{jk)}=0~
\label{G-anal}
\eea
and possesses the super-Weyl transformation  
\bea
\cG^{ij}&=&{\rm e}^{6\s} G^{ij}~.
\label{super-Weyl-G}
\eea

The explicit expression for $W_0$, eq. (\ref{W-BI-2}),  and the super-Weyl transformation law 
(\ref{super-Weyl-W}) imply 
\be
\cW_0 =1~.
\ee
Then, it follows from (\ref{Gij}) and (\ref{super-Weyl-G}) that 
\be
\cG_0^{ij}=-2 \cS^{ij} = {\rm e}^{6\s} G_0^{ij}~.
\ee

The supergeometry corresponding to the 5D $\cN=1$ anti-de Sitter superspace
is characterized by the following conditions  \cite{KT-Msugra3} (see also  \cite{KT-M}): 
\be
\cC_\ha{}^{ij}=0~,\qquad 
\cX_{\ha\hb} =\cN_{\ha\hb}=0~, \qquad \cS^{ij} \neq 0~.
\label{AdS-constraints}
\ee
Then, it follows from the Bianchi identities   \cite{KT-Msugra3} 
that $\cS^{ij} $ is covariantly constant, 
\be
\cD_\hal^{k} \cS^{ij} =0~.
\ee
As argued in \cite{BILS}, in the family of five-dimenisonal 
$\cN$-extended anti-de Sitter superspaces
\bea
{\rm AdS}^{5|8\cN}
= \frac{ {\rm SU}(2,2|\cN) }{ {\rm SO}(4,1) \times  {\rm U}(\cN)}~,
\non
\eea
it is only the case $\cN=1$ which corresponds to (locally)  conformally flat supergeometry
(although no explicit construction was given in \cite{BILS}). 
Below we will derive an explicit realization for the 5D $\cN=1$ anti-de Sitter superspace
as a locally conformally flat supergeometry.

Let us look for a supersymmetric extension of the ${\rm AdS}_5$ metric 
in Poincar\'e coordinates\footnote{These coordinates are known to cover 
one-half of the AdS hyperboloid.} 
\be
{\rm d}^2s = \Big(\frac{R}{z}\Big)^2 \Big( 
\eta_{mn}{\rm d}x^m   {\rm d}x^n +  {\rm d}z^2 \Big)~, \qquad 
R={\rm const}~, \qquad 
m =0, 1,2,3~
\ee
with $\eta_{mn}$ the four-dimensional Minkowski metric. 
The bosonic coordinates $x^m$ and $z$ are related to those
used in the main body of this paper as $x^{\hat m} = (x^m, z)$.
Since the supergeometry  $D_{\hat{A}} $ is flat, our first problem 
is  to look for a real superfield $W_0(z,\q^{\hat \m}_i)$ 
which solves  eq. (\ref{W-BI-2}) for the vector multiplet  
field strength in flat superspace.
There are at least three ways to address this problem: (i) brute-force approach; 
(ii) harmonic superspace construction; (iii) projective superspace construction. 
In the first case, one starts
with  a general superfield $W_0(z,\q^{\hat \m}_i)$ and then tries 
to satisfy eq. (\ref{W-BI-2}). In the second and third approaches, one starts with a useful 
ansatz for the harmonic or projective  prepotential for a 5D $\cN=1$ vector multiplet, 
and then read off the corresponding field strength following the  rules given in 
\cite{KL, K}. In all cases, it is convenient to express the four-component 
Grassmann coordinates, $\q^{\hat \a}_i$,  in terms of two-components 
spinors (see \cite{KL} for more details, including the explicit expressions for 
the 5D gamma-matrices in terms of the sigma-matrices etc.).
 \bea
\q^{\hat \a}_i = ( \q^\a_i , - {\bar \q}_{\dt \a i})~, 
\qquad
\q_{\hat \a}^i =   
\left(
\begin{array}{c}
\q_\a^i \\
{\bar \q}^{\dt \a i}    
\end{array}
\right)~, \qquad 
\overline{\q^\a_i} = {\bar \q}_{\dt \a}^{ i}    
\eea
as well as to express the 5D $\cN=1$ spinor covariant 
derivatives $D_{\hat \a}^i$ (without central charge)
in terms of 4D $\cN=2$  spinor covariant 
derivatives $D_{\a}^i$ and ${\bar D}_{\dt \a i}$  (with central charge)
following \cite{KL}
\bea
D^i_{\hat \a}  = \left(
\begin{array}{c}
D_\a^i \\
{\bar D}^{\dt \a i}    
\end{array}
\right)~, 
\qquad 
D^{\hat \a}_i  = 
(D^\a_i \,, \, -{\bar D}_{\dt \a i}) 
\label{con}
\eea 
where 
\bea
 D^i_\a &=& \frac{\pa}{\pa \q^{\a}_i}
+ {\rm i} \,(\s^b )_{\a \bd} \, {\bar \q}^{\dt \b i}\, \pa_b
+\q^i_\a \pa_z ~, 
\qquad
{\bar D}_{\dt \a i} 
= - \frac{\pa}{\pa {\bar \q}^{\dt \a i}} 
- {\rm i} \, \q^\b _i (\s^b )_{\b \dt \a} \,\pa_b
-{\bar \q}_{\dt \a i} \pa_z ~. 
\label{4D-N2covder1}
\eea
The most general expression for the field strength $W_0(z,\q^{\a}_i, {\bar \q}_{\dt \a}^j)$
can be shown to be:
\bea
W_0  &=&
A
+{\rm i}\big(\q_{ij} -\qb_{ij}\big) B^{ij}
-{1\over 12}\big( \q^4 +\qb^4 \big) \pa^2_zA
+{\rm i}\, \q^k{}_{(i}\qb_{j)k}\pa_zB^{ij}
+\hf\q_{ij}\qb^{ij}\pa^2_zA
\non\\
&&
+{{\rm i}\over 12}\big( \q^4\qb_{ij}- \q_{ij}\qb^4\big)\pa^2_zB^{ij}
+{1\over 144}\q^4\qb^4\pa^4_zA
~,~~~~~~~~~
\label{W-0}
\eea
where 
\be
\q_{ij} := \q^{\a}_i\q_{\a j}~, 
\qquad {\bar \q}^{ij} := {\bar \q}^i_{\dt \a } {\bar \q}^{\dt \a j}~, 
\qquad \overline { \q_{ij}} ={\bar \q}^{ij}~, 
\qquad \q^4 :=\q^{ij} \q_{ij} ~, \qquad 
{\bar \q}^4 := \overline{\q^4}~.
\ee
Here $A(z)$ and $B^{ij}(z) =B^{ji}(z) $ are real functions of $z$, 
\be
\overline{A}=A~, \qquad \overline{B^{ij}} =B_{ij}~, 
\ee
but otherwise are completely arbitrary.

With $W_0$ given as in  eq. (\ref{W-0}), 
we have satisfied the first constraint in (\ref{AdS-constraints}).
The next problem is to solve the second constraint in (\ref{AdS-constraints}), 
$\cX_{\ha\hb} =0$ or, equivalently, $\cN_{\ha\hb}=0$.
Its  solution is as follows:
\bea
A(z) = \frac{R}{z} ~, \qquad 
B^{ij}(z) =- \frac{R}{2z^2}\, 
{\bm s}^{ij}~, 
\qquad 
{\bm s}^{ij}:= \frac{s^{ij}}{ \sqrt{\hf s^{ij}s_{ij}} }~,
\label{AdSfieldst}
\eea
with 
\be
R={\rm const}~, 
\qquad s^{ij}=s^{ji}={\rm const}~, 
\qquad 
\overline{s^{ij}}=s_{ij}
~.
\ee
It is a short calculation to demonstrate that the {\it covariantly constant} torsion $\cS^{ij} $
is 
\be
\cS^{ij} = \frac{1}{R} \, 
{\bm s}^{ij}
~+~O(\q)~.
\ee
This completes our explicit realization of ${\rm AdS}^{5|8}$
as (locally) conformally flat superspace.

Let us leave ${\rm AdS}^{5|8}$ for a while, and 
discuss the structure of a manifestly supersymmetric action principle 
in the  case of an arbitrary conformally flat superspace.  
In accordance with the supergravity formulation developed in 
\cite{KT-Msugra3,KT-Msugra2}, 
the supersymmetric action is generated by 
a {\it covariant projective supermultiplet} of weight two, 
$\cL^{++}(u^+)$, which is defined to be holomorphic with respect to additional 
isotwistor variables
$u^+_i \in {\mathbb C}^2 \setminus\{0\}$.
The fact that the Lagrangian is projective and has weight $+2$, 
means the following:
\be
u^+_i \cD^i_{\hat \a} \cL^{++}(u^+) =0~, \qquad
 \cL^{++}(c \,u^+) = c^2 \, \cL^{++}(u^+) ~, \qquad 
 c\in {\mathbb C}  \setminus\{0\}~, 
 \ee
see   \cite{KT-Msugra3} for more details, including the reality condition
of $\cL^{++}$, $\widetilde{\cL}{}^{++}=\cL^{++}$, 
with respect to the so-called smile conjugation.
The action is
\bea
S(\cL^{++})&=&
\frac{1}{6\pi} \oint_C (u^+ \rd u^{+})
\int \rd^5 x \,{\rm d}^8\q\,\cE\, \frac{\cL^{++}}{(\cS^{++})^2}~,
\qquad \cE^{-1}={\rm Ber}\, (\cE_{\hat A}{}^{\hat M})~.
\label{InvarAcW=1}
\eea
Here  $C$ is a closed integration contour,
$\cS^{++}(u^+): =\cS^{ij} u^+_iu^+_j $ and  
$ (u^+ \rd u^{+}) := u^{+i} \rd u^{+}_i$.

Let us choose a coordinate system in which the covariant derivatives
$\cD_{\hat A}$ are related to the flat global ones, $D_{\hat A}$, 
according to eqs. (\ref{D_hal-ConfFlat}--\ref{D_a-ConfFlat-2}).
We then have 
\bea
\cE&=& \re^{-2\s} = W_0~, 
\qquad
-2 \cS^{++} = W_0^{-3} G_0^{++}~, 
\eea
with
\bea
G_0^{++}
: =G_0^{ij} u^+_iu^+_j
=\ri\,D^{\hal +}W_0D_\hal^{+} W_0+{\ri\over 2} W_0 D^{\hal +} D^+_{\hat \a} W_0~, 
\qquad D^+_{\hat \a} G_0^{++} =0
\eea
and $D^+_{\hat \a}:= D^i_{\hat \a} u^+_i$. 
We also have 
\bea
 \cL^{++} = W_0^{-3} L^{++}~, \qquad D^+_{\hat \a} L^{++} =0~.
 \eea
Here $L^{++}(u^+)$ is a {\it rigid  projective supermultiplet} of weight $+2$ 
living in flat 5D $\cN=1$ superspace ${\mathbb R}^{5|8}$.

More generally, if $\cQ^{(n)}(u^+)$ is a {\it covariant projective supermultiplet} of weight $n$, 
\be
u^+_i \cD^i_{\hat \a} \cQ^{(n)}(u^+) =0~, \qquad
 \cQ^{(n)}(c \,u^+) = c^n \, \cQ^{(n)}(u^+) ~, \qquad 
 c\in {\mathbb C}  \setminus\{0\}~, 
\ee
it is generated by a  {\it rigid  projective supermultiplet} of weight $n$, 
$Q^{(n)}(u^+) $, living in ${\mathbb R}^{5|8}$.
\bea
\cQ^{(n)} = W_0^{-3n/2} Q^{(n)}~, \qquad D^+_{\hat \a} Q^{(n)} =0~.
\label{curved-flat} 
\eea

The above action turns into\footnote{In general, 
the transformation (\ref{D_hal-ConfFlat}--\ref{D_a-ConfFlat-2})
relating the ``flat'' and  ``curved'' covariant derivatives, 
can be defined only locally, as in the case of ${\rm AdS}^{5|8}$. 
Although the locally supersymmetric action (\ref{InvarAcW=1}) is globally defined, 
its ``flat'' form (\ref{InvarAc2}) holds in general locally. 
In this paper, we do not discuss global issues.} 
\bea
S(\cL^{++})&=&
\frac{2}{3\pi} \oint_C (u^+ \rd u^{+})
\int \rd^5 x \,{\rm d}^8\q\, \frac{ L^{++}W_0^4}{(G_0^{++})^2}~.
\label{InvarAc2}
\eea
Using the identity  \cite{KT-Msugra3}
\bea
D^{(+4)}W_0^4
={3\over 4}(G_0^{++})^2~, \qquad 
(D^+)^4:=-{1\over 96}\ve^{\hal\hbe\hga\hde}
D^+_{\hal} D^+_{\hbe}D^+_{\hga}D^+_{\hde}~,
\label{idW}
\eea
we can next transform $S(\cL^{++})$ as follows:
\bea
S(\cL^{++})&=&\left.
\frac{2}{3\pi} \oint_C \frac{(u^+ \rd u^{+})}{(u^+u^-)^4}
\int \rd^5 x \,
(D^-)^4(D^+)^4\left\{ 
 \frac{ L^{++}W_0^4}{(G_0^{++})^2}\right\} \right|_{\q=0}    \non \\
 &=&\left.
\frac{1}{2\pi} \oint_C \frac{(u^+ \rd u^{+})}{(u^+u^-)^4}
\int \rd^5 x \,
(D^-)^4 L^{++}  \right|_{\q=0}~.
\label{InvarAc3}
 \eea
 Here 
\bea
(D^-)^4:=-{1\over 96}\ve^{\hal\hbe\hga\hde}
D^-_{\hal} D^-_{\hbe}D^-_{\hga}D^-_{\hde}~, \qquad
D^-_{\hat \a} := u^-_i D^i_{\hat \a}~,
\eea
and the isotwistor $u^-_i$ introduced 
is  constrained to obey 
the inequality $(u^+u^-) \neq 0$ 
(which means that  $u^+_i$ and $u^-_i$ are linearly independent)
but otherwise is completely arbitrary.

It is possible to transform the action further and represent it as an integral 
over 4D $\cN=1$ superspace \cite{K,KT-M}. First of all, we note that 
the action is invariant under arbitrary
projective transformations of the form
\be
(u_i{}^-\,,\,u_i{}^+)~\to~(u_i{}^-\,,\, u_i{}^+ )\,R~,~~~~~~R\,=\,
\left(\begin{array}{cc}a~&0\\ b~&c~\end{array}\right)\,\in\,{\rm GL(2,\mathbb{C})}~.
\label{projectiveGaugeVar}
\ee
This   symmetry implies that the action is actually independent of $u^-_i$, 
and that the isotwistor $u^+_i$ provides homogeneous coordinates for ${\mathbb C}P^1$. 
Second, without loss of generality, we can assume that the integration contour 
$C$  does not intersect the north pole of  ${\mathbb C}P^1$.  
We thus can chose 
\bea
u^{+i} &=&u^{+\1}(1,\z) 
\equiv u^{+\1}\z^i ~,\qquad 
u^-_i =(1,0) ~, 
\label{zeta}
\eea
as well as 
\bea
L^{++}(u^+) =\ri(u^{+\1})^2\z \,L(\z)~, 
\eea
with $\z$ the complex local coordinate parametrizing ${\mathbb C}P^1$.  
Now, the constraint $D^+_{\hat \a} L^{++} =0$ is equivalent to 
$\z^i D_{i \hat \a } L(\z)=0$. The latter can be used to rewrite (\ref{InvarAc3})
in the form: 
\bea
S(\cL^{++})&=&
\frac{1}{2\pi \rm i} \oint_C  \frac{\rd \z}{\z}
\int \rd^5 x \,\rd^4 \q \,
L(\z) \Big|_{\q^\a_{\2}=0}~.
\label{InvarAc4}
\eea
In this form, the supersymmetric action is given in terms of $\cN=1$ 
superfields.\footnote{Eq. (\ref{InvarAc4}) is the 5D $\cN=1$ version of the projective 
superspace action principle \cite{KLR,LR}.}

If the Lagrangian $L^{++}$ is independent of the vector multiplet 
associated with $W_0$, 
then the action (\ref{InvarAc3}) contains no information  about the curved supergeometry, 
and thus (\ref{InvarAc3})
 describes a rigid superconformal theory of  the general type studied in \cite{K}. 
 An example of such theories is the general superconformal nonlinear sigma-model
 formulated in terms  of  
{\it covariant arctic weight-one} multiplets 
$\U^{+ } (u^+) $ and their smile-conjugates
$ \widetilde{\U}^{+}$   and described by the Lagrangian
\cite{K,KT-M,K2}
\bea
\cL^{++} = {\rm i} \, K(\U^+, \widetilde{\U}^+)~,
\label{conformal-sm}
\eea
with $K(\F^I, {\bar \F}^{\bar J}) $ a real analytic function
of $n$ complex variables $\F^I$, where $I=1,\dots, n$.
${}$For $\cL^{++}$ to be a weight-two real projective superfield, 
it is sufficient to  require 
\bea
 \F^I \frac{\pa}{\pa \F^I} K(\F, \bar \F) =  K( \F,   \bar \F)~.
 \label{Kkahler2}
 \eea

Let us  give an example of 
dynamical systems
with the Lagrangian $L^{++}$ depending on the vector multiplet $W_0$.
Following \cite{KT-M,KT-Msugra1}, 
consider the  system of interacting {\it covariant arctic weight-zero} multiplets 
${\bf \U}(u^+)$   and their smile-conjugates
$ \widetilde{ \bf{\U}}$  described by the Lagrangian 
\bea
\cL^{++} = \hf \cS^{++}\,
{\bf K}({\bf \U}, \widetilde{\bf \U})~,
\eea
with ${\bf K}(\F^I, {\bar \F}^{\bar J}) $ a real function 
which is not required to obey any 
homogeneity condition. 
${}$For this model,
the line integral in (\ref{InvarAc4}) should be carried out around the origin.
Because ${\bf \U}(u^+)$ has vanishing weight, $n=0$, 
eq.  (\ref{curved-flat}) means that ${\bf \U}(u^+) = {\bf \U}(\z) $ 
is a rigid projective supermultiplet. 
The corresponding flat-superspace form of the Lagrangian is 
\bea
L^{++} = -\frac{1}{4} G_0^{++}\,
{\bf K}({\bf \U}, \widetilde{\bf \U})~.
\label{model}
\eea
The action can be seen to be  invariant under K\"ahler transformations of the form
\be
{\bf K}({\bf \U}, \widetilde{\bf \U})~\to ~{\bf K}({\bf \U}, \widetilde{\bf \U})
+{\bf \L}({\bf \U}) +{\bar {\bf \L}} (\widetilde{\bf \U} )~,
\ee
with ${\bf \L}(\F^I)$ a holomorphic function.

To describe the dynamics of Yang-Mills supermultiplets, 
we should introduce a gauge field $V_0(u^+)$ for the Abelian vector 
multiplet $W_0$ associated with our conformally flat superspace. 
The $V_0(u^+)$ is a {\it tropical weight-zero} multiplet such that 
the field strength is given as \cite{K}
\be
W_0  = \frac{1}{ 16\pi {\rm i}} \oint
\frac{ (u^+{\rm d} u)}{(u^+ u^-)^2}  \,
({D}^-)^2  \, V_0(u^+)~, \qquad ({D}^-)^2:=D^{-\hat \a}D^-_{\hat \a}~.
\label{strengt4}
\ee
Since $V_0$ has vanishing weight, $n=0$, 
eq.  (\ref{curved-flat}) means that $V_0(u^+) = V_0(\z) $ 
is invariant under the super-Weyl transformations, 
i.e. $\cV_0=V_0$. 
The field strength $W_0$ is invariant under the gauge transformations
\be
V_0 ~\to ~ V_0 + \l +\widetilde{\l}~,
\label{gauge}
\ee
with $\l(u^+) $ an arbitrary {\it arctic weight-zero} superfield.
Let $\cW$ be the gauge-covariant field strength of a Yang-Mills 
supermultiplet, and $\cV(u^+)$ is a gauge field
(i.e. a tropical weight-zero multiplet taking its values in the Lie algebra
of the gauge group). 
Then, we can construct the covariant projective weight-two multiplet
\be
\cG^{++}(u^+):=\cG^{ij} u^+_i u^+_j~,
\ee
with $\cG^{ij} $ given in (\ref{Gij}). Dynamics of the Yang-Mills 
supermultiplet can be described by the Lagrangian
\be
\cL^{++}_{\rm YM} = \frac{1}{g^2}\, \cV_0 \, \cG^{++} + \k \,\cG^{++}_0 \,{\rm tr}\cV~,
\label{YM}
\ee  
with $g$ and $\k$ the coupling constants. The corresponding action 
can be seen to be invariant under the gauge transformations
(\ref{gauge}). 
The second term in (\ref{YM}) is a Fayet-Iliopoulos term. 

If the K\"ahler potential ${\bf K}(\F^I, {\bar \F}^{\bar J}) $ in 
(\ref{model}) corresponds to a K\"ahler manifold with isometries, 
on can gauge the sigma-model following \cite{K3}.
In particular, one can generate ``massive'' sigma-models
if the gauging is carried out using the frozen vector multiplet 
 $V_0(\z)$.

As follows from (\ref{model}), all information about the 
curved superspace geometry is now encoded in 
$ G_0^{++}(u^+)= G_0^{ij} \, u^+_iu^+_j$. 
In the case of the anti-de Sitter superspace ${\rm AdS}^{5|8}$, 
this superfield can be shown to be
\bea
G_0^{++}(u^+) &=&
-{2 R^2 \over  z_c^3} \Big\{ 
{\bm s}^{++}
-{3\ri  \over z_c} 
\Big((\q^{+})^2 - (\qb^{+})^2\Big)
-{3  \over  z_c (u^+u^-)}
\Big( (\q^{+})^2 +(\qb^{+})^2 \Big) {\bm s}^{+-}
\non\\
&&
+{12  \over  z_c^2(u^+u^-)^2} 
(\q^{+})^2 (\qb^{+})^2 {\bm s}^{--}\Big\} ~.
\label{G_0^{++}}
\eea
Here ${\bm s}^{\pm \pm }= {\bm s}^{ij} u^\pm_i u^\pm_j$, 
\be
z_c =z -\frac{1}{(u^+u^-)} \Big( \q^+ \q^- + {\bar \q}^+{\bar \q}^-\Big)~, 
\ee
and $\q^\pm_\a = \q^i_\a u^\pm_i$ and ${\bar \q}^\pm_{\dt \a} = {\bar \q}^i_{\dt \a} u^\pm_i$.
The variables $z_c$, $\q^+_\a$ and ${\bar \q}^+_{\dt \a}$, which appear in  
the right-hand side of (\ref{G_0^{++}}), are  annihilated by $D^+_{\hat \a}$, 
that is, they are analytic in the sense of 
the 5D $\cN=1$ version \cite{KL} of the 
harmonic superspace approach \cite{GIKOS,GIOS}.
One can check that $G^{++}_0$ is independent of $u^-$, 
\be
\frac{\pa }{\pa u^-} G_0^{++}=0~, 
\ee
in spite of the fact that separate contributions
to the right-hand side of (\ref{G_0^{++}}) do depend on $u^-$.

Let us now represent $G_0^{++}(u^+)$, eq. (\ref{G_0^{++}}),  as
\bea
G_0^{++}(u^+)=\ri(u^{+\1})^2\z \,G_0(\z)~.
\eea
Instead of giving the complete expression for $G_0(\z)$,  
it is sufficient to consider $G_0(\z)$ in the limit of   $\q_\2^\a=\qb^\2_{\dt \a}=0$, 
since only this truncated expression for $G_0(\z)$ appears in the action 
(\ref{InvarAc4}).
Defining
\bea
\q^\a:=\q^\a_\1~,\qquad
\qb_{\dt \a}:=\q_{\dt \a}^\1~, 
\eea
a short calculation gives
\bea
G_0(\z)|_{\q_\2=0}&=&
{2\ri R^2  \over z^3}\Big\{ \Big(
\z {\bm s}^{\1\1}-2{\bm s}^{\1\2}+{1\over\z}{\bm s}^{\2\2}\Big)
+{3\over  z}\q^2\Big( {\bm s}^{\1\1}-{1\over\z}( {\bm s}^{\1\2}+\ri)\Big)
\non\\
&&
+{3 \over  z}\qb^2\Big(- {\bm s}^{\2\2}+\z( {\bm s}^{\1\2}+\ri)\Big)
+{12 \over  z^2}\q^2\qb^2(
{\bm s}^{\1\2}+\ri) \Big\}
~.
\label{G-02}
\eea
${}$For completeness, we also give the expression for $W_0$ 
in the limit of   $\q_\2^\a=\qb^\2_\ad=0$.
\bea
W_0|_{\q_\2=0}&=&
{R\over  z}
-{\ri R \over2 z^2} \Big( \q^2 {\bm s}^{\1\1}
-\qb^2{\bm s}^{\2\2} \Big)
-{\ri R\over z^3} \, \qb^2\q^2({\bm s}^{\1\2}+\ri)
~.
\label{W-0-reduced}
\eea

Up to an SU(2) rotation, one can always choose ${\bm s}^{ij}$ to 
have the form:
\be
{\bm s}^{\1\1}={\bm s}^{\2\2}=0 \qquad \Longleftrightarrow \qquad
{\bm s}^{\1\2}=\pm \ri~.
\label{S11=0}
\ee
Now, it follows from (\ref{G-02}) and (\ref{W-0-reduced})
\bea
{\bm s}^{\1\2}=- \ri \qquad \Longrightarrow \qquad
W_0|_{\q_\2=0}=
{R\over  z}~, \qquad 
G_0(\z)|_{\q_\2=0}=- 
{4 R^2  \over z^3} ~.
\label{G-03}
\eea
It is seen that the superfields $W_0|_{\q_\2=0}$
and  $G_0(\z)|_{\q_\2=0}$ are invariant under 
the standard 4D $\cN=1$ super-Poincar\'e transformations.  

It is not difficult to see that 
the second solution, ${\bm s}^{\1\2}= \ri $, in eq. (\ref{S11=0})
 simply corresponds to the replacement
$(\q^\a_{\1}, {\bar \q}^{\1}_{\dt \a}) \to (\q^\a_{\2}, {\bar \q}^{\2}_{\dt \a})$
in the above consideration. 
In particular, we have 
\bea
{\bm s}^{\1\2}= \ri \qquad \Longrightarrow \qquad
G_0(\z)|_{\q_\1=0}= {4 R^2  \over z^3} ~.
\label{G-04}
\eea

With the choice (\ref{G-03}), the action (\ref{InvarAc4})
generated by (\ref{model}) becomes 
\bea
S &=&
\frac{1}{R} \oint_C  \frac{\rd \z}{2\pi {\rm i} \z}
\int \rd^5 x \,\rd^4 \q \,
\Big(\frac{R}{z} \Big)^3\,
{\bf K}({\bf \U}, \widetilde{\bf \U}) ~.
\label{InvarAc5}
\eea
Here the dynamical variables are 
\bea
{\bf \U} (\z) &=& \sum_{n=0}^{\infty} {\bf \U}_n  \,\z^n
= \F +\z \S +\dots~, 
\qquad
\widetilde{\bf \U} (\z) = \sum_{n=0}^{\infty} 
\frac{ (-1)^n}{\z^n}  {\bar {\bf \U}}_n \,
= {\bar \F} -\frac{1}{\z} {\bar \S} +\dots ~, ~~~~
\label{pm}
\eea
where the two leading components of ${\bf \U} (\z) $ are constrained 
4D $\cN=1$ superfields, 
\be 
{\bar D}^{\dt \a} \, \F =0~,
\qquad
-{1\over 4} {\bar D}^2 \, \S 
= \pa_z\,  \F~. 
\label{pm-constraints}
\ee 
The other components of $\U (\z)$ are complex unconstrained superfields, 
and they appear to be non-dynamical (auxiliary) in the model under consideration.

In the free case, 
\be
{\bf K}({\bf \U}, \widetilde{\bf \U}) = R\, \widetilde{\bf \U} \,\bf \U ~, 
\label{free}
\ee
one can easily do the contour integral in (\ref{InvarAc5}) to result with 
\bea
S &=&
\int \rd^5 x \,\rd^4 \q \,
\Big(\frac{R}{z} \Big)^3\,\Big( {\bar \F}\F - {\bar \S}\S \Big) ~+~
\dots
\eea
where the omitted terms involve the auxiliary superfields. 
The latter  terms vanish on the equations of motion for the auxiliary
superfields.
The quadratic action obtained can be shown to agree 
(upon implementing a superfield Legendre transformation
that converts $\S$ into a chiral superfield)
with the model previously constructed
in  \cite{MP} (see also \cite{CMT})  by rewriting 
supersymmetric component actions  in AdS${}_5$ in 
terms of 4D $\cN=1$ superfields.

Since the explicit $z$-dependence in (\ref{InvarAc5})
is not accompanied  by any $\z$-dependence, 
the auxiliary superfields can be eliminated in the   AdS${}_5$
case in the same  way it has been done in the flat global case
for a large class of nonlinear sigma-models, 
see e.g. \cite{AKL}.

To describe off-shell massive hypermultiplets living in AdS${}^{5|8}$, 
it is necessary to have at our disposal
a gauge field $V_0(\z)$ that generates the corresponding 
field strength $W_0$. 
Assuming the SU(2) choice (\ref{S11=0}), one can check that 
$V_0(\z)$ can be chosen to be
\bea
V_0(\z) ={R\over z_c{\z}}\Big({\bm\q}^2(\z)-{\bm\qb}^2(\z)\Big)
+{\ri R\over z^2_c\z^2}{\bm\q}^2(\z){\bm\qb}^2(\z){\bm s}^{\1\2}
~,
\label{V_0-proj}
\eea
where
\bea
{\bm\q}^{\a}(\z)&=&-\z\q^{\a}_\2-\q^\a_\1~,\qquad
{\bm\qb}_{\dt \a}(\z)=-\z\qb_{\dt \a}^{\1}+\qb_{\dt \a}^{\2}~, \non \\
z_c&=&z+(\q_{\1\2}-\qb^{\1\2})+\z(\q_{\2\2}+\qb^{\1\1})~.
\eea
The corresponding field strength (\ref{strengt4}) 
can be checked to agree with (\ref{AdSfieldst}).
Projecting to the 4D $\cN=1$ superfields gives 
\bea
V_0|_{\q_\2=0}=
{R\over  z}\Big({1\over\z}\q^2-\z\qb^2\Big)
+{\ri R\over  z^2}\q^2\qb^2({\bm s}^{\1\2}+\ri)
~,
\label{V_0-proj-N=1}
\eea
and therefore
\bea
{\bm s}^{\1\2}=- \ri \qquad \Longrightarrow \qquad
V_0|_{\q_\2=0}=
{R\over  z} \Big(  {1\over\z}\q^2-\z\qb^2\Big)
~.
\eea
The massive hypermultiplet Lagrangian is obtained
by replacing (\ref{free}) with  
\be
{\bf K}({\bf \U}, \widetilde{\bf \U}, V_0) = R\, \widetilde{\bf \U} \, {\rm e}^{m V_0}\bf \U ~, 
\ee
with $m $ the hypermultiplet mass. 
This model is invariant under gauge transformations
\be
V_0 ~\to ~ V_0 + \l +\widetilde{\l}~, \qquad 
 \U ~\to ~ {\rm e}^{-m\l } \U ~,
\label{gauge2}
\ee
with the gauge parameter $\l(\z)$ an arctic superfield. 
In conclusion, we note that the prepotential (\ref{V_0-proj}) 
should be used in the Lagrangian  (\ref{YM}) to 
describe the dynamics  
of  the  Yang-Mills  supermultiplet  in AdS${}^{5|8}$.
\\

\noindent
{\bf Acknowledgements:}\\
One of us (SMK) 
acknowledges useful discussions with 
Jonathan Bagger and Dmitry Belyaev 
about conformally flat superspaces. 
This work is supported  in part by the Australian Research Council.


\small{

}


\begin{thebibliography}{66}

\bibitem{RS}
  L.~Randall and R.~Sundrum,
  ``A large mass hierarchy from a small extra dimension,''
  Phys.\ Rev.\ Lett.\  {\bf 83}, 3370 (1999)
  [hep-ph/9905221].

\bibitem{ABN}
  R.~Altendorfer, J.~Bagger and D.~Nemeschansky,
  ``Supersymmetric Randall-Sundrum scenario,''
  Phys.\ Rev.\  D {\bf 63}, 125025 (2001)
  [hep-th/0003117].

\bibitem{GP}
  T.~Gherghetta and A.~Pomarol,
  ``Bulk fields and supersymmetry in a slice of AdS,''
  Nucl.\ Phys.\  B {\bf 586}, 141 (2000)
  [hep-ph/0003129].

\bibitem{FLP}
  A.~Falkowski, Z.~Lalak and S.~Pokorski,
  ``Supersymmetrizing branes with bulk in five-dimensional supergravity,''
  Phys.\ Lett.\  B {\bf 491}, 172 (2000)
  [hep-th/0004093].

\bibitem{BK}
I.~L. Buchbinder and S.~M. Kuzenko, {\it Ideas and Methods of Supersymmetry and
Supergravity, Or a Walk Through Superspace}, IOP, Bristol, 1998.

\bibitem{Ohashi}
  T.~Kugo and K.~Ohashi,
  ``Off-shell d = 5 supergravity coupled to matter-Yang-Mills system,''
  Prog.\ Theor.\ Phys.\  {\bf 105}, 323 (2001)
   {[hep-ph/0010288]};
  T.~Fujita and K.~Ohashi,
  ``Superconformal tensor calculus in five dimensions,''
  Prog.\ Theor.\ Phys.\  {\bf 106}, 221 (2001)
 {[hep-th/0104130]}.

\bibitem{Bergshoeff}
 E.~Bergshoeff, S.~Cucu, M.~Derix, T.~de Wit, R.~Halbersma and A.~Van Proeyen,
  ``Weyl multiplets of N = 2 conformal supergravity in five dimensions,''
  JHEP {\bf 0106}, 051 (2001)
  [hep-th/0104113].


\bibitem{KT-Msugra3}
  S.~M.~Kuzenko and G.~Tartaglino-Mazzucchelli,
  ``Super-Weyl invariance in 5D supergravity,''
  arXiv:0802.3953 [hep-th].

\bibitem{KT-Msugra1}
S.~M.~Kuzenko and G.~Tartaglino-Mazzucchelli,
  ``Five-dimensional superfield supergravity,''
    Phys.\ Lett.\ B {\bf 661}, 42 (2008),
 [arXiv:0710.3440].

\bibitem{KT-Msugra2}
  S.~M.~Kuzenko and G.~Tartaglino-Mazzucchelli,
  ``5D supergravity and projective superspace,''
  JHEP {\bf 0802}, 004 (2008) [arXiv:0712.3102].


\bi{KT-M}
S.~M. Kuzenko and G. Tartaglino-Mazzucchelli,
``Five-dimensional N=1 AdS superspace:
Geometry,  off-shell multiplets and dynamics,''
Nucl. Phys. B {\bf 785}, 34 (2007),
[arXiv:0704.1185].
  

\bibitem{BILS}
  I.~A.~Bandos, E.~Ivanov, J.~Lukierski and D.~Sorokin,
  ``On the superconformal flatness of AdS superspaces,''
  JHEP {\bf 0206}, 040 (2002)
  [arXiv:hep-th/0205104].

  
\bibitem{KL}
  S.~M.~Kuzenko and W.~D.~Linch, III,
  ``On five-dimensional superspaces,''
  JHEP {\bf 0602}, 038 (2006)
  [hep-th/0507176].

\bibitem{K}
 S.~M.~Kuzenko,
 ``On compactified harmonic/projective superspace, 5D superconformal
 theories, and all that,''
  Nucl.\ Phys.\  B {\bf 745}, 176 (2006)
  [hep-th/0601177]. 
  
  \bibitem{K2}
S.~M.~Kuzenko, ``On superconformal projective hypermultiplets,''
JHEP {\bf 0712}, 010 (2007) [arXiv:0710.1479].

\bibitem{KLR}
A. Karlhede, U. Lindstr\"om and M. Ro\v cek,
``Self-interacting tensor multiplets in N=2 superspace,''
Phys.\ Lett.\ B {\bf 147}, 297 (1984).

\bibitem{LR}
  U.~Lindstr\"om and M.~Ro\v{c}ek,
 ``New hyperk\"ahler  metrics  and new supermultiplets,''
  Commun.\ Math.\ Phys.\  {\bf 115}, 21 (1988);
  ``N=2 super Yang-Mills theory in projective superspace,''
  Commun.\ Math.\ Phys.\  
  {\bf 128}, 191 (1990).
  
 \bibitem{GIKOS}
A.~S.~Galperin, E.~A.~Ivanov, S.~N.~Kalitsyn, V.~Ogievetsky, E.~Sokatchev, 
``Unconstrained N=2 matter, Yang-Mills and supergravity theories in harmonic
superspace,''
Class.\ Quant.\ Grav.\  {\bf 1}, 469 (1984).
 
\bibitem{GIOS}
A.~S.~Galperin, E.~A.~Ivanov, V.~I.~Ogievetsky and E.~S.~Sokatchev,
{\it Harmonic Superspace}, Cambridge University Press,  2001.
  
\bibitem{K3}
  S.~M.~Kuzenko,
  ``On superpotentials for nonlinear sigma-models with eight supercharges,''
  Phys.\ Lett.\  B {\bf 638}, 288 (2006)
  [hep-th/0602050].
  
\bibitem{MP}
  D.~Marti and A.~Pomarol,
  ``Supersymmetric theories with compact extra dimensions in N = 1
  superfields,''
  Phys.\ Rev.\  D {\bf 64}, 105025 (2001)
  [hep-th/0106256].
  
\bibitem{CMT}
  G.~Cacciapaglia, G.~Marandella and J.~Terning,
  ``Dimensions of supersymmetric operators from AdS/CFT,''
  arXiv:0802.2946 [hep-th].
  
  \bibitem{AKL}
  M.~Arai, S.~M.~Kuzenko and U.~Lindstr\"om,
  ``Hyperkaehler sigma models on cotangent bundles of Hermitian symmetric
  spaces using projective superspace,''
  JHEP {\bf 0702}, 100 (2007)
  [hep-th/0612174]; 
  ``Polar supermultiplets, Hermitian symmetric spaces and hyperkahler
  metrics,''
  JHEP {\bf 0712}, 008 (2007) [arXiv:0709.2633 [hep-th]].
  

\end{thebibliography}
\end{document}